\documentclass[aps,amsfonts,nofootinbib,article,twocolumn,showpacs,superscriptaddress]{revtex4}
\usepackage{graphicx}
\usepackage{amsmath}
\usepackage{amsfonts}
\usepackage{amssymb}
\usepackage{gensymb}

\def\openone{\leavevmode\hbox{\small1\kern-3.8pt\normalsize1}}
\def\N{\leavevmode\hbox{ Z \kern-8 pt\normalsize{Z}}}
\def\openone{\leavevmode\hbox{\small1\kern-3.8pt\normalsize1}}
\def\openJ{\leavevmode\hbox{J \kern-9.5pt\normalsize J}}
\def\openS{\leavevmode\hbox{ S \kern-9.3pt\normalsize S}}
\newcommand{\bb}{\begin{equation}}
\newcommand{\ee}{\end{equation}}
\newcommand{\eqb}{\begin{eqnarray}}
\newcommand{\eqf}{\end{eqnarray}}

\usepackage{color}

\begin{document}

\title{Classical and Quantum Dispersion Relations}

\author{Sergio A. Hojman}
\email{sergio.hojman@uai.cl}
\affiliation{Departamento de Ciencias, Facultad de Artes Liberales,
Universidad Adolfo Ib\'a\~nez, Santiago 7491169, Chile.}
\affiliation{Centro de Investigaci\'on en Matem\'aticas, A.C., Unidad M\'erida, Yuc. 97302, M\'exico}
\affiliation{Departamento de F\'{\i}sica, Facultad de Ciencias, Universidad de Chile,
Santiago 7800003, Chile.}
\affiliation{Centro de Recursos Educativos Avanzados,
CREA, Santiago 7500018, Chile.}
\author{Felipe A. Asenjo}
\email{felipe.asenjo@uai.cl}
\affiliation{Facultad de Ingenier\'ia y Ciencias,
Universidad Adolfo Ib\'a\~nez, Santiago 7491169, Chile.}

\begin{abstract}
It is showed that, in general, classical and quantum dispersion relations are different due to the presence of the Bohm potential. There are exact particular solutions of the quantum (wave) theory which obey the classical dispersion relation, but they differ in the general case. The dispersion relations may also coincide when additional assumptions are made, such as WKB or eikonal approximations, for instance. This general result also holds for non--quantum wave equations derived from classical counterparts, such as in ray and wave optics, for instance. Explicit examples are given for covariant scalar, vectorial  and tensorial fields in flat and curved spacetimes.
\end{abstract}


\maketitle

\section{Introduction}

Dispersion relations have a long and illustrious history throughout different physics subjects ranging from fluid dynamics to particle physics, including Maxwell's electromagnetism and Einstein's gravity, among many others.  In the early days, at the beginning of the twentieth century, pioneers such as Wood and Raleigh discovered anomalous dispersion relations \cite{wood1,wood2,ray1,ray2} enriching the subject. Later on, the discovery of the connection between dispersion relations and causality, which opened a whole new area of research, was turned into one of the most striking results in the area (a rigorous teatment is found in \cite{toll}). There are many articles and books written on the subject of dispersion relations (see for example Refs.~\cite{kro,kra,pick,mesh,mar,fastlight}).

Furthermore, for over fifty years there have been both theoretical \cite{slep,skro,pleb,dewitt,vz1,mas1,har,mas2,mas3,sahthesis,hor,sah1,berry,riv1,ah0,ah1,ah2,petrov} and (more recently) experimental \cite{mugnai,gio,bouch,riv2,konda} results which seem to indicate that light propagation may also occur with either super-- or sub--luminal speeds in vacuum flat spacetime or on curved spacetimes as well as on dielectric media. The wavevectors of these solutions do not proceed along null geodesics. These non--geodesic solutions appear in addition to the usual ones where light propagation occurs along null geodesics in any media and in any kind of spacetimes.
Needless to say, if these results were firmly confirmed, experimental and/or observational works based on (Special or General) Relativity would have to be reconsidered and many results related to Astrophysics and Cosmology should be also in need of reexamination.

This work exhibits the relation that exists between non-vanishing Bohm potentials and superluminal (as well as subluminal) dispersion relations.
The super-- and sub--luminal character of the propagation may be stated in terms of (non--relativistically invariant) phase or group velocities but it may also be cast in the language of (relativistically invariant) dispersion relations. Customarily, the usual propagation of waves with the speed of light (written in terms of the wave four--vector $k_\mu$, the gradient of the wave's phase) is tantamount to
\begin{equation}
k_\mu k^\mu = 0\ ,   \label{kk1}
\end{equation}
while either super-- or sub--luminal wave propagation can be indicated as
\begin{equation}
k_\mu k^\mu \ne 0\ .  \label{kk2}
\end{equation}
The ``right hand side'' of Eq. \eqref{kk2} is negative (positive) for super-- (sub--)luminal propagation with the (West Coast) metric signature convention adopted here. 
Here, the spatial part of the
 wave four--vector $k_\mu$ allows us to measure the phase velocity of the  wave.

In general, a super--luminal propagation of fields is not a behavior forbidden by relativity. In fact, super--luminal propagation is a well-known effect occuring through anomalous dispersion in media (see Ref.~\cite{fastlight}, and references therein).
The reason for no violation with special theory of relativity is due that in those cases, the super--luminal group or phases velocities
are not carrying information, which is described by the signal velocity. This was already noted by
Sommerfeld and Brillouin \cite{fastlight}, by indicating  that different velocities can be associated to wave motion, as group velocity, phase velocity, signal velocity, energy transfer velocity, etc. and they are all different in absorbing media.

The problem, then, arises when waves are studied in vacuum, as one should not expect to have the aforementioned issues which take place in media. 
Below, we show that it is possible to have wave solutions with dispersion relations presenting a non--null geodesic behavior, even in vaccum. This occurs due to the variations of wave amplitude, that
affect the propagation of the wave, implying a non--local effect in the waves' trayectories. It is proved that this effect is contained in the so--called Bohm potential, which is always associated to any wave-like structure, even propagating in vaccum.  
The Bohm potential $V_B$ modifies the dispersion relations always in the non--traditional form
$k_\mu k^\mu=V_B$, changing  the phase velocity of the wave. This is what produces the super-- (sub--)luminal propagation of  massless scalar, vectorial and tensorial fields. 
In other words, the non--traditional form of dispersion relations for the phase velocities of massless fields, emerge as a general consequence of its wave-like nature (not being a point--particle description), and not only because it is travelling in a medium. In general, it is only when the wave dynamics is approximated to the eikonal limit, i.e. point particle-like nature, that $V_B\rightarrow 0$, and the massless field moves in null geodesics. 

There are multiple examples of systems whose classical (particle--like) and quantum (wave--like) dispersion relations are different, as it can be seen in \cite{slep,skro,pleb,dewitt,vz1,mas1,har,mas2,mas3,sahthesis,hor,sah1,berry,riv1,ah0,ah1,ah2,petrov}, for instance. 
These differences can also be tracked to Hamilton--Jacobi formalisms for classical and quantum theories. This allows us to connect these non--traditional dispersion relations with the foundations of quantum mechanics \cite{madelung,bohm,takab,book,book2}. 
These ideas have been pushed forward to study, for example, systems with identical classical and quantum dynamics \cite{makow}, with exact two--dimensional quantum solutions \cite{hadual}, or with the Bohm potential as an internal energy \cite{glendenis}.

It is the aim of this work to show that the Bohm potential plays the important role of differentiating  classical and quantum dispersion relations, 
bringing out the  wave--like structures of the fields. It is this extended property of the field, with non--local interactions due to a
 non--vanishing Bohm potential, that allows the field to have super-- or sub--luminal behavior. In order to explicitly show that, we present examples of this behavior for massless Klein--Gordon scalar fields, vectorial Maxwell fields and tensorial gravitational wave fields.

\section{Classical and Quantum Hamilton--Jacobi equations}

Consider classical mechanics, where a relation between energy $E$ and momentum $\vec{p}$, for point particles, may be established (usually through energy conservation). The quantization process translates this relation into the Schr\"odinger wave equation whose dispersion relation between energy $E$ (frequency $\omega$) and momentun ${\vec{p}}$ (wave vector ${\vec{k}}$) is different, in general, from the original one.
From the Lagrangian $L=({1}/{2}) m {\dot{\vec{r}}}^{2}-V$ of a classical point particle under a potential $V=V(\vec r)$,
 the 
 Hamilton--Jacobi (HJ) equation can be derived
\begin{equation}
\frac{1}{2m}\vec {\nabla} S \cdot \vec {\nabla} S + V(\vec{r})+\frac{\partial S(\vec{r},t)}{\partial t}=0, \label{HJ}
\end{equation}
where
$S$ represents the classical action. The HJ equation can be considered as the dispersion relation of the classical system.

Let us now consider the quantum theory for such particle. This is described by  complex wavefunctions $\psi$ and $\psi^*$ satisfying the  Schr\"odinger equations
\begin{eqnarray}
\left[-\frac{{\hbar}^2}{2m}\nabla^2 + V(\vec{r}) - i \hbar \frac{\partial}{\partial t}  \right] \psi (\vec{r},t) &=& 0\, ,\label{sch1}\\
\left[-\frac{{\hbar}^2}{2m}\nabla^2 + V(\vec{r}) + i \hbar \frac{\partial}{\partial t}  \right] \psi^*(\vec{r},t) &=& 0\, \label{sch2}, 
\end{eqnarray}
for one particle moving in the presence of a (real, time independent) potential $V(\vec{r})$.
Now let us rewrite the complex wavefunction $\psi$ in polar form
\begin{equation}
\psi = A \exp\left(\frac{i S}{\hbar}\right)\, , \label{polar}   
\end{equation}
where $A=A(\vec{r},t)$ and $S=S(\vec{r},t)$ are real functions. Notice that the information of the quantum system is encoded in $A$ and $S$ in the same way that is in $\psi$ and $\psi^*$.

Then the Schr\"odinger equations may be written as a pair of nonlinear coupled real equations \cite{book,book2}
\begin{eqnarray}
\frac{1}{2m}\vec {\nabla} S \cdot \vec {\nabla} S - \frac{\hbar^2}{2m}\frac{ {\nabla}^2 A }{A} + V +\frac{\partial S}{\partial t}&=&0\, , \label{HJB1} \\  \frac{1}{m} \vec{\nabla} \cdot (A^2 \vec{\nabla} S)  +\frac{\partial A^2}{\partial t}&=&0\, . \label{cont}    
\end{eqnarray}
Eq.~\eqref{HJB1} define the Quantum Hamilton--Jacobi (QHJ) equation \cite{book2}. It differs from the classical HJ equation \eqref{HJ},  for a particle moving on a potential $V(\vec{r})$, by the addition of the Bohm potential for a non--relativistic particle
$V_B$ which is defined by
\begin{equation}
V_B=-\frac{\hbar^2}{2m}\frac{ {\nabla}^2 A }{A}\, , \label{VB}    
\end{equation}
\noindent where $A=\sqrt{\psi^*\psi}$ is the amplitude of the wavefunction $\psi$. 
Furthermore,  it is  added to the system the probability conservation (continuity) equation  \eqref{cont} to the theory.
Both equations \eqref{HJB1} and \eqref{cont} are known as the Madelung--Bohm (MB) equations \cite{madelung,bohm}.  These constitute the hydrodynamical version of quantum mechanics.
Note that $\hbar$ appears in the Bohm potential only.

It is clear that in the quantum theory, the MB  system and the QHJ equation is a (non--trivial) modification of the HJ equation due to the presence of the Bohm potential. So, in general, the classical and quantum dispersion relations are not equivalent except in the cases for which the Bohm potential vanishes.
For the case of plane waves (and other particular solutions) the Bohm potential vanishes identically.  In the two dimensional case, if $A(x,y)$ is a harmonic function of the coordinates, the Bohm potential also vanishes \cite{hadual}. On the other hand, in the WKB approximation and eikonal approximations, the Bohm potential is neglected assuming slowly varying wave amplitudes.

As an opposite case,  Berry and Balasz \cite{berry} described  a wave packet solution to the {\it {free}} one--dimensional particle Schr\"odinger equation (written in terms of an Airy function) that propagates without distortion and  non--vanishing time dependent {\it {acceleration}} in spite of the {\it {absence of a force}}. The Berry--Balasz is a solution that produces  a non--vanishing Bohm potential, which may be consider as the origin of such phenomenon.

Considering the previous discussion, we study the massless cases in order to find the effect of Bohm potential on dispersion relations for the field propagation. We present different cases for massless scalar,  vectorial and tensorial fields in which their MB associated equations are different from a HJ theory.
We show how subluminal and superluminal solutions emerges as solutions for non--vanishing Bohm potential of the respective equations, where the luminal (lightlike) behavior occurs when the Bohm potential vanishes (and the equation coincides with a HJ theory).

\section{The complex wave equation in flat-spacetime}

Consider the covariant wave equation \cite{ah3} 
\begin{equation}
\Box u  =0, \label{WE}
\end{equation}
for for a massless  (complex) scalar field function, where $\Box\equiv\partial_\mu\partial^\mu$ is the flat-spacetime d'Alembert operator in any coordinates [with signature $(-,+,+,+)$]. Let us write
$u=U\ e^{iS}$ for real functions $U$ and $S$.
Then, Eq.~\eqref{WE} separates in
\begin{eqnarray}
k_\mu k^\mu&=&V_B=\frac{\Box U}{U}\, \label{WE2a}\\
\partial_\mu\left(k^\mu U^2\right)&=&0\, .\label{WE2b}
\end{eqnarray}
In Eq.~\eqref{WE2a}, $k_\mu=\partial_\mu S$ is the wavevector for the propagation of the field. The above equations
contains the Bohm potential  $V_B$ of a scalar field,
which is, in general, nonzero. Eq.~\eqref{WE2a} is the dispersion relation, which
is equal to the Bohm potential. This is the equivalent to the QHJ equation for a free massless particle. Besides,
Eq.~\eqref{WE2b} is the continuity equation for the scalar field. 

As long as $V_B= 0$, the field has null geodesic (luminal) propagation. This also occurs in the eikonal limit.
In such cases, the dispersion relation is $k_\mu k^\mu=0$. This corresponds to the simplest solution which may be constructed with a constant amplitude, such as a plane wave with form $u(t,x)=  \exp\left[i k(x-ct)\right]$, with constant $k$.  The massless scalar equation coincides with a theory described by a HJ equation only when $V_B=0$.

However, if $V_B\neq 0$, other kind of behaviors are possible. 
A very explicit propagating solution with nonzero Bohm potential was found by Slepian in 1949 \cite{slep}, in cartesian coordinates $(t,x,y)$. These solutions have the form $u_{>,<}(t,x,y)= U_{>,<}(y) \exp\left[i k(x-vt)\right]$, where
\begin{eqnarray}
U_>(y)&=& A_> \cos \left(k y \sqrt{\frac{v^2}{c^2}-1} \right) \ ,  \label{u>}\\
U_<(y)&=& A_< \cosh \left(k y \sqrt{1-\frac{v^2}{c^2}} \right) \ , \label{u<}
\end{eqnarray}
for constant $k$,   $c$, $A_>$ and $A_<$. 
Here, $v$ is the constant phase velocity of the wave.
Solution \eqref{u>} is valid for super--luminal phase velocities $v>c$. However, its dispersion relation is
\begin{eqnarray}
\left.k_\mu k^\mu\right|_>=V_{B>}=k^2 \left( 1-\frac{v^2}{c^2}\right) < 0\,,
\label{VB>}
\end{eqnarray}
and such propagation is timelike (sub--luminal).
On the contrary, solution \eqref{u<} is only valid for sub--luminal phase velocities, $v<c$, but the dispersion 
relation for this solution 
\begin{eqnarray}
\left.k_\mu k^\mu\right|_<=V_{B<}=  k^2 \left({\frac{v^2}{c^2}-1}\right) > 0\, ,  \label{VB<}
\end{eqnarray}
is space--like, and then the wave propagates at super--luminal speed.
This kind of behavior (that phase velocities and dispersion relations hint in opposite ways) is a common phenomenon for  waves in a medium. But, we stress that the above solutions are in vacuum.

In both above case, Bohm potentials $V_{B>}$ and $V_{B<}$ do not vanish (except for the case $v=c$).
Thus, these constant  negative and positive Bohm potentials correspond to super--luminal and sub--luminal propagation.
Of course, plane wave (and other particular) solutions have vanishing Bohm potentials (and therefore, the ``right hand sides'' of the dispersion equations are zero).

In general, different luminality character of the solutions may be traced back to a non--vanishing Bohm potential  or modified forms of it. This is due to their dynamics is described by QHJ equation. The same can occur in  vectorial equations, such as Maxwell  for instance, as we show below.

\section{Maxwell equations and its Bohm potential}

Maxwell equations can be written in terms of the electromagnetic vector potential $A^\rho (x^\alpha)$ on a flat or curved spacetime background described (in general) by the metric $g_{\mu \nu} (x^\beta)$. For a vectorial equation, Bohm potential is a more complicated function, that does not depend only on the field amplitude but also on its vectorial features, such as polarization.

Below, we analyze how the Bohm potential affects electromagnetic propagation in flat and curved spacetime.

\subsection{In vacuum and flat spacetime}

The covariant form of Maxwell equations in flat spacetime are $\partial_\mu F^{\mu\nu}=0$. Writing the electromagnetic tensor $F_{\mu\nu}=\partial_\mu A_\nu-\partial_\nu A_\mu$ in terms of the electromagnetic potential, Maxwell equations becomes simply
\begin{equation}\label{eMax1flat}
\partial_\mu\left(\partial^\mu A^\nu-\partial^\nu A^\mu\right)=0\, .
\end{equation}
Under the Lorenz gauge
$\partial_\mu A^\mu=0$,
the above equation reads
\begin{equation}\label{eMax2flat}
\Box A^\nu=0\, .
\end{equation}
Then, Maxwell equations are reduced to complex wave equation \eqref{WE} 
for each polarization. This implies that the complete analysis of previous section applies for each polarization of the electromagnetic field, having a general non--zero Bohm potential.
Considering  the form \cite{ah2}
\begin{equation}
A_\mu=\xi_\mu e^{iS}   \label{ansa}
\end{equation}
for the potential, 
where $\xi_\mu$ (amplitude of the electromagnetic field) and $S$ (its phase) are real functions of the spacetime coordinates, then Maxwell equation \eqref{eMax2flat} separates as
\begin{eqnarray}
\xi^\nu\left(k_\mu k^\mu\right)&=&\Box\xi^\nu\, , \label{disp1Mxflat}\\
\partial_\mu\left(k^\mu \xi^\nu\right)+k^\mu\partial_\mu\xi^\nu&=&0\, .\label{contMaxflat}
\end{eqnarray}
where $k_\mu  = {\partial}_\mu S$ is the wavevector. Eq.~\eqref{disp1Mxflat} gives rise to the dispersion relation
\begin{equation}
k_\mu k^\mu=V_B=\frac{\xi_\nu\Box\xi^\nu}{\xi^2}\, , \label{disp1Mxflat2}
\end{equation}
where $\xi\equiv\sqrt{\xi_\mu\xi^\mu}$. Notice that now the Bohm potential takes into account the spacetime variations of the amplitude and polarization of the wave. Anew, Eq.~\eqref{disp1Mxflat2} has the role of the QHJ equation for the electromagnetic wave.

On the other hand, Eq.~\eqref{contMaxflat} produces the continuity equation for photon propagation
\begin{equation}
\partial_\mu\left(k^\mu \xi^2\right)=0\, .\label{WE2b}
\end{equation}
Finally, Lorenz gauge reduces to 
\begin{equation}
    \partial_\mu \xi^\mu=0\, ,\qquad \xi^\mu k_\mu=0\, .\label{LorGaugeMax}
\end{equation}

The simplest solution for electromagnetic waves are plane waves with constant amplitude, such that $V_B=0$ and null geodesic behavior $k_\mu k^\mu=0$. However, because of electromagnetic fields satisfies Eq.~\eqref{eMax2flat}, then Slepian \cite{slep} solutions  \eqref{u>} and \eqref{u<} are also solutions for electromagnetic waves propagating at super-- or sub--luminal velocities in vacuum  with non--vanishing Bohm potential. Consider a particular solution, polarized in a $z$--direction for instance, following the Slepian ansatz. In that case, we have $A_z(t,x,y)= \xi_z(y)\exp\left(i kx-ik vt\right)$, where  $v>c$ is the superluminal phase velocity of the electromagnetic wave, and
\begin{eqnarray}
\xi_z(y)&=&\xi_0\cos\left(ky\sqrt{\frac{v^2}{c^2}-1}\right)\, ,\\
k_0&=& -{k v}\, ,\quad k_x=k\, \quad k_y=0=k_z\, . 
\end{eqnarray}
This solutions solves Eqs.~\eqref{disp1Mxflat} and \eqref{contMaxflat}, with a constant Bohm potential
\eqref{VB>}, and a timelike (subluminal) dipersion relation 
\begin{equation}
k_\mu k^\mu=k^2\left(1-\frac{v^2}{c^2}\right)<0\, .
\end{equation}
 Also, it fulfills  Lorenz gauge \eqref{LorGaugeMax}. This solution determines that the electromagnetic field can travel sub--luminally in vacuum. 
This behavior, that is typical for electromagnetic plane waves propagating in a medium, now it is obtained in vacuum.
Of course, superluminal solution are straightforward to be obtained to be
\begin{eqnarray}
\xi_z(y)=\xi_0\cosh\left(ky\sqrt{1-\frac{v^2}{c^2}}\right)\, .
\end{eqnarray}

The above solutions for sub--luminal and super--luminal electromagnetic waves correspond to the superposition of two plane waves (each of them propagating in null geodesics). However, their sum can travel at $v\neq c$, phenomenon called the {\it scissor effect}. Therefore, the above solutions represent real electromagnetic waves.

\subsection{In vacuum and curved spacetime}

Maxwell equations
in curved spacetime are $\nabla_\alpha F^{\alpha\beta}=0$, where $\nabla_\mu$ is the covariant derivative on a curved spacetime background. Written in terms of the electromagnetic vector potential $A^\rho (x^\alpha)$, we get
\begin{equation}\label{eMax1}
\partial_\alpha\left[\sqrt{-g}g^{\alpha\mu}g^{\beta\nu} (\partial_\mu A_\nu-\partial_\nu A_\mu)\right]=0\, ,
\end{equation}
 with the metric $g_{\mu \nu} (x^\beta)$ [and where where $g^{\alpha\beta}$ is the inverse of the metric $g_{\mu \nu}$ and $g$ is its determinant].

In general \cite{ah2}, for the form \eqref{ansa}, Maxwell equation separates into the two following equations
\begin{eqnarray}
\left(k^\mu k_\mu\right) \xi^\beta&=&\frac{1}{\sqrt{-g}}\partial_\alpha\left[\sqrt{-g}g^{\alpha\mu}g^{\beta\nu}\left(\partial_\mu \xi_\nu-\partial_\nu\xi_\mu\right)\right]\, ,\label{SeparadaMax1}\\
0&=&\partial_\alpha\left[\sqrt{-g}g^{\alpha\mu}g^{\beta\nu}\left(k_\mu \xi_\nu-k_\nu\xi_\mu\right)\right] \nonumber\\ & & + \sqrt{-g}k^\mu g^{\beta\nu}\left(\partial_\mu \xi_\nu-\partial_\nu\xi_\mu\right)\, ,\label{SeparadaMax2}
\end{eqnarray}
where now $k_\mu \equiv {\nabla}_\mu S  = {\partial}_\mu S$.
Eq.~\eqref{SeparadaMax1} gives rise to the dispersion relation
\begin{equation}\label{SeparadaMax3}
k^\mu k_\mu =V_B\equiv\frac{\xi_\beta}{\sqrt{-g}\, \xi^2}\partial_\alpha\left[\sqrt{-g}g^{\alpha\mu}g^{\beta\nu}\left(\partial_\mu \xi_\nu-\partial_\nu\xi_\mu\right)\right]\, .
\end{equation}
This generalized antisymmetric Bohm potential $V_B$ for vector fields contains now information of the polarization of the fields and the curvature of the spacetime. Eq.~\eqref{SeparadaMax3} becomes the curved spacetime analogue version of the QHJ equation for electromagnetic propagation.
Other authors \cite{riv1} have found equivalent results where the non--null geodesic behaviour of light waves is associated to a non--vanishing Bohm potential.

On the other hand, Eq.~\eqref{SeparadaMax2} produces the photon conservation in curved spacetimes \cite{ah2}
\begin{equation}
\nabla_\mu(\xi^2 k^\mu)=0\, .
\end{equation}

The above Maxwell equations are more complicated in curved spacetime. Again, plane waves solutions (with constant amplitude) are solution of Eqs.~\eqref{SeparadaMax1} and \eqref{SeparadaMax2}, with $V_B=0$, and defining a null geodesic behavior for light in curved spacetime $k_\mu k^\mu=0$.
Nevertheless,
simple Slepain solutions can be found for some non--trivial metrics. This solutions have non--vanishing Bohm potential, and thus shows non--null geodesic behavior. For example, in a flat cosmological model, with the  metric $g_{\mu\nu}=(-1,a^2,a^2,a^2)$ in cartesian coordinates, and with $a=a(t)$, it can be shown that we can find a subluminal solution for electromagnetic potential in a $z$--direction. This solution reads
\begin{equation}
    A_z(t,x,y)=\xi_0\cos\left(ky\sqrt{\frac{v^2}{c^2}-1}\right)\exp\left(ikx-i{k v}\int\frac{dt}{a}\right)\, ,
\end{equation}
and it solves Eqs.~\eqref{SeparadaMax1}
and \eqref{SeparadaMax2}, for super--luminal phase velocity $v>c$. In this case, this solution has non--zero  and time--dependent Bohm potential
\begin{equation}
    k_\mu k^\mu=V_B=\frac{k^2}{a^2}\left(1-\frac{v^2}{c^2}\right)<0\, ,
\end{equation}
and therefore it represents sub--luminal (timelike) light propagating in a  cosmological universe.
It also solves  
the Lorenz gauge ${\nabla}_\mu A^\mu = 0$ in curved spacetime, which translate into the equations $k_\mu\xi^\mu = 0$ and $\nabla_\mu \xi^\mu=0$. A super--luminal solution  for light in a cosmological background
is 
\begin{equation}
A_z(t,x,y)=\xi_0\cosh\left(ky\sqrt{1-\frac{v^2}{c^2}}\right)\exp\left(ikx-i{k v}\int\frac{dt}{a}\right)\, , 
\end{equation}
only valid for $v<c$.
These electromagnetic waves can only propagates at speed of light when $v=c$, and thus they have constant amplitude and vanishing Bohm potential, i.e., they are plane waves.

\section{Gravitational waves and their Bohm potential}

It is clear that the Bohm potential introduces non--local effects due to the extended properties of the fields, producing its non--geodesic behavior. Therefore, it is expected to find a Bohm potential for the propagation of gravitational waves. In a vaccum curved background spacetime, gravitational waves are described by the equation \cite{misner}
\begin{equation}
\nabla_\alpha\nabla^\alpha {\bar h}_{\mu\nu}+2R^{(B)}_{\mu\alpha\nu\beta}{\bar h}^{\alpha\beta}=0\, ,
\label{gravwabes}
\end{equation}
where ${\bar h}_{\mu\nu}=h_{\mu\nu}-(1/2) h\, g_{\mu\nu}^{(B)}$, where $g_{\mu\nu}^{(B)}$
is the background metric (that produces the vaccum curved spacetime) and $h_{\mu\nu}$ is the perturbed metric, such that the total metric is given by $g_{\mu\nu}=g_{\mu\nu}^{(B)}+h_{\mu\nu}$. Thus, $h=g_{\mu\nu}^{(B)}h^{\mu\nu}$, and 
$R^{(B)}_{\mu\alpha\nu\beta}=g_{\mu\gamma}^{(B)}{R^{(B)\gamma}}_{\alpha\nu\beta}$
is the Riemann tensor associated to the curved background metric. Therefore, the covariant derivative $\nabla_\mu$ is taken in the background metric.
Wave eq.~\eqref{gravwabes} must be complemented by the Lorentz gauge condition
\begin{equation}
\nabla^\alpha {\bar h}_{\alpha\mu}=0\, ,
\label{gravwabes2}
\end{equation}
in order to ensure the  two dynamic degree of freedom of the gravitational field.

Let us consider a gravitational wave with the form
\begin{equation}
{\bar h}_{\mu\nu}=\zeta_{\mu\nu} e^{iS}\, ,
\end{equation}
as before, but now for a tensor field. It is straightforward to show that Eq.~\eqref{gravwabes} produces two equations. The first one is
\begin{equation}
\left(k_\alpha k^\alpha\right) \zeta_{\mu\nu}=\nabla^\alpha \nabla_\alpha\zeta_{\mu\nu}+2R^{(B)}_{\mu\alpha\nu\beta} \zeta^{\alpha\beta}\, ,
\label{Bohmpotential1}
\end{equation}
where now $k_\mu=\nabla_\mu S$ is the four-wavevector of the gravitational wave. Also, we get the equation
\begin{equation}
\nabla_\alpha\left(k^\alpha \zeta_{\mu\nu}\right)+k_\alpha \nabla^\alpha \zeta_{\mu\nu}=0\, .
\label{conser1gravwaves}
\end{equation}
On the other hand, the gauge condition \eqref{gravwabes2} gives
\begin{eqnarray}
\nabla^\alpha\zeta_{\alpha\mu}=0\, ,\quad k^\alpha \zeta_{\alpha\mu}=0\, .
\end{eqnarray}

Anew, in general, a gravitational wave cannot travel in null geodesics, due to the existence of Bohm potential $V_B$ for gravitational waves. From Eq.~\eqref{Bohmpotential1} we obtain that
\begin{eqnarray}
k_\alpha k^\alpha&=&V_B\nonumber\\
&=& \frac{1}{\zeta^2}\left(\zeta^{\mu\nu}\nabla^\alpha \nabla_\alpha\zeta_{\mu\nu}+2 R^{(B)}_{\mu\alpha\nu\beta}\zeta^{\mu\nu}\zeta^{\alpha\beta}\right)\, ,
\label{Bohmpotential2}
\end{eqnarray}
where $\zeta=\sqrt{\zeta^{\alpha\beta}\zeta_{\alpha\beta}}$. Furthermore, from Eq.~\eqref{conser1gravwaves}, we obtain the graviton conservation 
\begin{equation}
\nabla_\alpha\left(k^\alpha \zeta^2\right)=0\, .
\label{conser2gravwaves}
\end{equation}

The above Bohm potential $V_B$ for gravitational waves is different from Bohm potential for vector and scalar fields of previous sections. However, it retains the same kind of features of  dependence on variations of the wave amplitude.
Also,  Bohm potential \eqref{Bohmpotential2} shows the coupling between the amplitude wave with spacetime curvature, but it also exists in a flat--spacetime
$V_B\rightarrow \zeta^{\mu\nu}\Box\zeta_{\mu\nu}/{\zeta^2}$, indicating that the null geodesic behavior of gravitational waves is  achievable for constant amplitude, or in eikonal--limit of the waves. 

This implies that even in an empty  flat--spacetime, and in analogue form to the studied previous cases for Klein--Gordon and Maxwell fields,
 gravitational waves can have phase velocities different from speed of light, allowing them to have a time--like behavior. An example of this is the solution $h_{zz}(t,x,y)=\xi_{zz}(y)\exp(i kx-ikvt)$, with phase velocity $v>c$, and 
$\xi_{zz}(y)=\cos\left(ky\sqrt{v^2/c^2-1}\right)$, that satisfies the wave equation and Lorentz gauge condition. 
 For this solution
\begin{equation}
k_\alpha k^\alpha=V_B=k^2\left(1-\frac{v^2}{c^2}\right)<0\, ,
\end{equation}
and the gravitational wave follow time--like trayectories, that are not geodesics.

\section{Summary, Conclusions and Outlook}

We have showed that quantum (or wave) differential equations give rise to dispersion relations that, in general, differ from those exhibited by classical (or particle) counterparts. 
We have been able to trace, quite in general, that those non--traditional dispersion relations are linked to a non--vanishing Bohm potential, which in turn means that they are associated to important variations of the wave amplitude. This is a consequence of the extension of the wave (as opposed to the point--like character of particles), and it is linked to the difference between the HJ and the QHJ equations.

 We have showed that Bohm potential modifies the dispersion relations of massless Klein-Gordon field. However, these results are more striking for light.
The non--traditional dispersion relations give rise to super--and sub--luminal propagation 
of light waves \cite{slep,skro,pleb,dewitt,vz1,mas1,har,mas2,mas3,sahthesis,hor,sah1,berry,riv1,ah0,ah1,ah2,petrov}, which seem to have been detected experimentally \cite{mugnai,gio,bouch,riv2,konda}.
The effect of that kind of light propagation, if confirmed, would tremendously impact both theoretical, experimental and observational
work in Relativity, Optics, Astrophysics and Cosmology. 
Furthermore, analogue resuts are obtained for gravitational waves, displaying how Bohm potential change their dispersion relation, modifying its geodesic behavior. This is relevant, for example, to determine the arrival time of a gravitational wave that has experienced gravitational lensing \cite{takaryu,Suyama}.

Of course, these non--traditional dispersion relations are present in any kind of dynamics that have both a classical (pointlike) and quantum (wavelike) propagation versions, regardless of the fact that particles (or fields) involved are massless or massive.
We would like to point out that in the references listed there are two kinds of non--traditional dispersion relations. One that we would like to call geometrical or kinematical is associated with the findings of Slepian \cite{slep} and Bohm potential, for instance, where one has just a wave packet that does the trick, and a physical or dynamical kind where the super-- or sub--luminal behavior of light may be traced back to the interaction of light polarization and metric rotation (or anisotropy) \cite{ah2,ah3}.
We would finally like to point out that, other authors \cite{ac1,mag1}, in order to get non--traditional ($k_\mu k^\mu \neq 0$) dispersion relations have postulated different kinds of models with modifications (or violation) of Lorentz invariance in order to accommodate experimental results.

It is perhaps interesting to remark that good old--fashioned perfectly generally covariant (or Lorentz covariant in the flat spacetime case) theories as the ones discussed here, give rise to non--traditional dispersion relations without the need of introducing exotic models, which may mean that at least some experimental data may be obtained without modifying Lorentz covariance.

\end{document}